\pdfoutput=1 

\documentclass{article}

\usepackage{amsmath}
\usepackage{mathtools}

\usepackage{biblatex} %
\usepackage{graphicx}

\usepackage{hyperref}
\usepackage[capitalise, noabbrev]{cleveref}

\usepackage[affil-it]{authblk}

\usepackage{caption}
\usepackage{subcaption}

\newsavebox\affbox

\providecommand{\keywords}[1]
{ %
    \vspace{.25cm}
    { \small \textbf{\textit{Keywords ---}} #1}
}

\title{Do Segmentation Models Understand Vascular Structure? A Blob-Based XAI Framework}

\author[1]{Guillaume Garret}
\author[1]{Antoine Vacavant}
\author[2]{Carole Frindel}

\affil[1]{Université Clermont Auvergne, CHU Clermont-Ferrand, Clermont Auvergne INP, CNRS, Institut Pascal, F-63000 Clermont-Ferrand, France} 
\affil[2]{Université Lyon I, Université Claude Bernard, Laboratoire CREATIS, CNRS, Inserm, INSA-Lyon, \par Lyon, France}

\date{}

\graphicspath{ {./resources/} } %

\begin{document}

\maketitle

\section{Abstract}
    Deep learning models have achieved impressive performance in medical image segmentation, yet their black-box nature limits clinical adoption.
In vascular applications, trustworthy segmentation should rely on both local image cues and global anatomical structures, such as vessel connectivity or branching.
However, the extent to which models leverage such global context remains unclear.
We present a novel explainability pipeline for 3D vessel segmentation, combining gradient-based attribution with graph-guided point selection and a blob-based analysis of Saliency maps.
Using vascular graphs extracted from ground truth, we define anatomically meaningful points of interest (POIs) and assess the contribution of input voxels via Saliency maps.
These are analyzed at both global and local scales using a custom blob detector.
Applied to IRCAD and Bullitt datasets, our analysis shows that model decisions are dominated by highly localized attribution blobs centered near POIs.
Attribution features show little correlation with vessel-level properties such as thickness, tubularity, or connectivity --- suggesting limited use of global anatomical reasoning.
Our results underline the importance of structured explainability tools and highlight the current limitations of segmentation models in capturing global vascular context.

\keywords{explainable artificial intelligence \and attribution \and vessel segmentation}

\section{Introduction}
    Reconstruction of vascular anatomical structures and detection of adjacent lesions are essential steps in clinical protocols, but their automation remains a constant challenge \cite{Goni_2022_BrainVesselSegmentationUsingDeepLearningReview, Moccia_2018_BloodVesselSegmentationAlgorithmsReview}.
Nowadays, the unmissable methods for meeting this need are based on deep learning.
To slightly improve the quality of vascular segmentation, the community has to design increasingly complex models, using ensembling \cite{Isensee_2021_nnUNetASelfConfiguringMethodForDeepLearningBasedBiomedicalImageSegmentation, Survarachakan_2021_EffectsOfEnhancementOnDeepLearningBasedHepaticVesselSegmentation}, attention modules \cite{Sun_2020_SAUNetShapeAttentiveUNetForInterpretableMedicalImageSegmentation, Oktay_2018_AttentionUNetLearningWhereToLookForThePancreas, Yan_2021_AttentionGuidedDeepNeuralNetworkWithMultiScaleFeatureFusionForLiverVesselSegmentation} or joint-models \cite{Zhang_2020_GraphAttentionNetworkBasedPruningForReconstructing3DLiverVesselMorphologyFromContrastedCTImages, Shin_2019_DeepVesselSegmentationByLerningGraphicalConnectivity}.

However, deep learning models might be seen as black-boxes.
While the number of parameters and non-linearities increase, the models become progressively more difficult to explain and interpret.
This lack of transparency poses a significant barrier for their adoption in high-stakes fields like healthcare.
In a recent publication, Chouvarda \emph{et al.} \cite{Chouvarda_2025_DifferencesInTechnicalAndClinicalPerspectiveOnAIValidationInCancerImagingMindTheGap} reported AI explainability and interpretability as the main concerns for clinical researchers.
Ensuring that the basis of models' decisions aligns with domain expert knowledge is then crucial.
In particular, a reliable vessel segmentation should rely on both local features, like shape and intensity, as well as global features, such as vessel inter-connections, which are meaningful in a broader context.

To address the opacity of deep models, the AI community has turned its attention to eXplainable Artificial Intelligence (XAI).
In this relatively new area of research, attribution-based explanations are among the most popular approaches.
Ranging from simple methods using gradients~\cite{Ancona_2018_TowardsBetterUnderstandingOfGradientBasedAttributionMethods} to more robust methods inspired by cooperative game theory~\cite{Anconna_2019_ExplainingDeepNeuralNetworksWithPolinomialTimeAlgorithmForShapleyValueApproximation, Lundberg_2017_UnifiedApproachToInterpretingModelPredictions}, attribution aims to identify the most influential input features, \emph{e.g.} voxels, in the model's decision-making process.
In this context, Rajpurkar \emph{et al.} \cite{Rajpurkar_2017_CheXNetRadiologistLevelPneumoniaDetectionOnChestXRaysWithDeepLearning} use CAM \cite{Zhou_2016_LearningDeepFeaturesForDiscriminativeLocalization} to ensure the classifier makes its decision focusing on the right pathological areas.
Using Integrated Gradients \cite{Sundararajan_2017_AxiomaticAttributionForDeepNetworks}, the authors of \cite{Wargnier-Dauchelle_2021_MoreInterpretableClassifierForMultipleSclerosis} leverage attribution to compare two training strategies for multiple sclerosis detection.
Finally, in \cite{Wargnier-Dauchelle_2023_WeaklySupervisedGradientAttributionConstraintForInterpretableClassificationAndAnomalyDetection}, a pathology classifier is constrained with a new loss function that expects attribution to be negative on healthy areas.

Although attribution methods are commonly used to explain classification models, their application to segmentation tasks is less frequent.
Indeed, as segmentation inherently involves localization, attribution maps are often considered redundant in this context. 
Additionally, many attribution methods are logit-specific, rather than class-specific, making the interpretation of attribution more challenging.
A few attempts have been made to explain medical segmentation models.
For instance Saleem \emph{et al.} \cite{Saleem_2021_VisualInterpretabilityIn3DBrainTumorSegmentationNetwork} identify the impact of multiple MRI sequences on model decisions using per-channel explanations derived from a CAM-based approach.
In their study, \cite{Hasany_2023_SegXResCAMExplainingSpatiallyLocalRegionsInImageSegmentation} introduce Seg-XRes-CAM, a method that generates layer-level and class-discriminative attribution for liver and stomach segmentation.

In our work, we argue that while segmentation is inherently a spatial computer task, attribution can still provide valuable insights into how and why a model is making its decisions, potentially leading to improved model performance and interpretability.
Considering segmentation task as pixel-wise classification, our main contribution consists in the design of a novel XAI pipeline dedicated to the interpretation of vessel segmentation models.
On two vascular datasets, we extract ground-truth graphs and define anatomically meaningful points of interest (POIs), such as bifurcations, endpoints, and mid-branches.
We compute gradient-based attribution maps at these POIs to highlight the input regions most influential to the model’s predictions.
To analyze the structure of these influential regions, we introduce a blob-based approach that detects compact attribution zones.
By comparing blob characteristics to vessel features like connectivity, thickness, and tubularity, we aim to quantify the impact of these high-level features on the decision-making process.

\section{Methodology}
\begin{figure}[!ht]
    \centering
    \includegraphics[width=\textwidth]{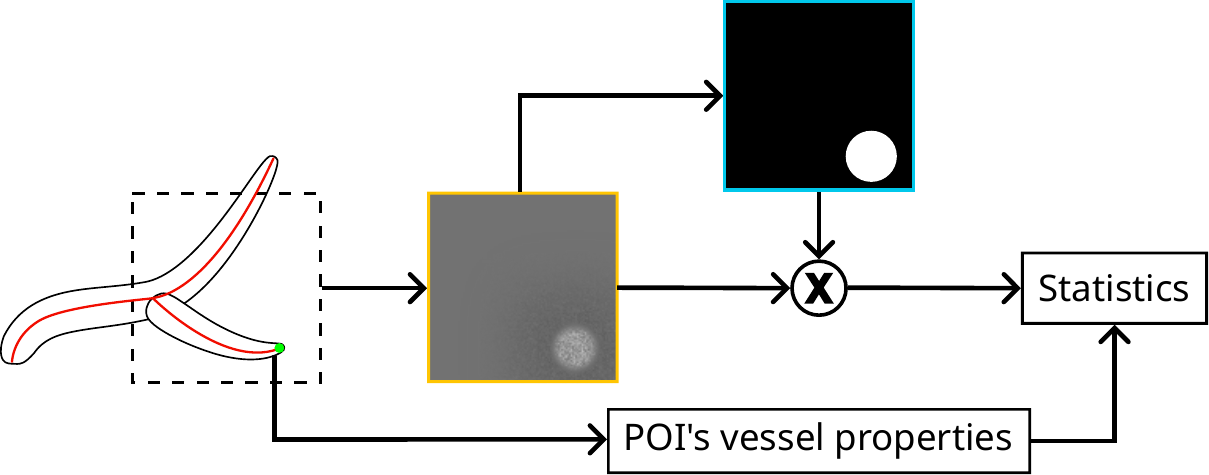}
    \caption{The proposed XAI pipeline. The red curves represent the vascular graph, with the green point indicating the selected POI. The resulting Saliency map (orange) is processed to compute the blob mask (cyan) and analyzed using our blob-based approach. Finally, the Saliency characteristics are combined with the POI's vessel properties to assess their contribution to the model's decision.}
    \label{fig:xai_pipeline}
\end{figure}

    This section details the components of our explainability framework, from the extraction of vascular points of interest to the computation and analysis of Saliency maps. 
The diagram presented in \cref{fig:xai_pipeline} depicts the key stages of the pipeline.

\subsection{Vascular Points of Interest} \label{sec:vascularPOIs}
In the following, we will use "voxel" to refer to an input voxel and "logit" to refer to an output voxel of the model.

Gradient-based attribution methods are logit-specific, meaning they are not directly aware of the classes.
In the context of classification, each class is represented by its own logit, where the value is $1$ for the predicted class and $0$ elsewhere.
Hence, the predicted class is a natural abstraction and corresponds to the label associated to the logit with the highest value.
In contrast, label-encoded segmentation does not have a single logit associated with a class, as the output and input have the same dimension.
In this latter context, every output logits can belong to any class depending on its value, meaning that we can not derive the "vessel class" all at once.
Instead, we must consider segmentation as a voxel-wise classification, leading to attribution relative to a single voxel.
However, computing attribution for every independant logit seems not relevant and is computationally heavy.
Therefore, it is essential to identify points of interest --- specific logits --- that we aim to explain.

\begin{figure}[!ht]
    \centering
    \includegraphics[width=0.45\textwidth]{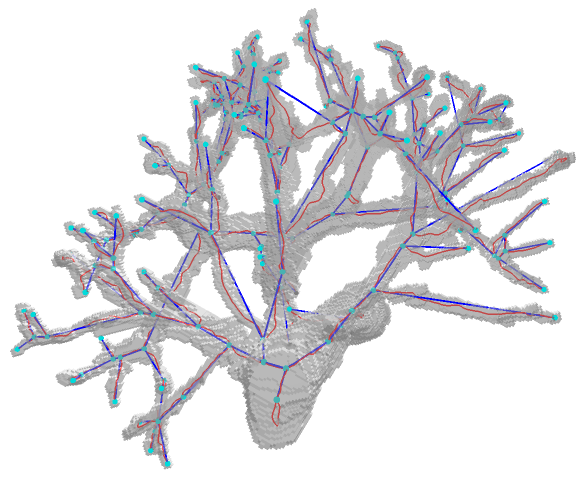}
    \includegraphics[width=0.45\textwidth]{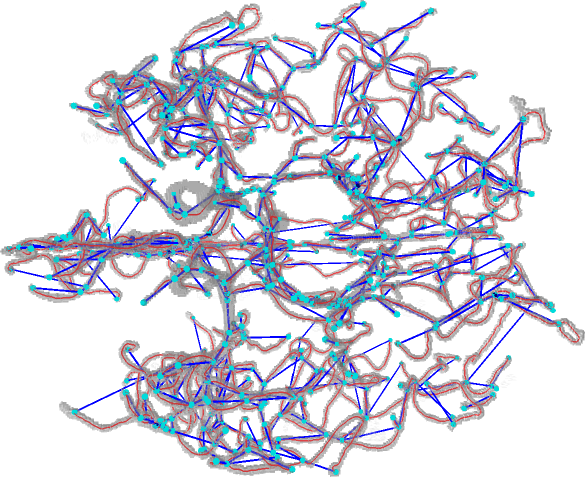}
    \caption{Images produced using \href{https://www.voreen.uni-muenster.de}{Voreen}. Vascular structure graphs from IRCAD (left) and Bullitt (right) samples. The vessel ground-truth is shown in grey, with cyan dots representing the graph nodes and red curves the indicating the centerlines.}
    \label{fig:voreen_graph_samples}
\end{figure}

Vascular networks are tree-like structures that can be depicted by a hierarchical graph.
As vessel recognition should partly be based on global context, \emph{e.g.} connections to other vessels, we produce explanations according to vascular structure.
Given the ground-truth of the vessels, we extract the vasculature using the vessel graph extraction pipeline from the Voreen framework~\cite{Drees_2021_ScalableRobustGraphAndFeatureExtractionForArbitraryVesselNetworksInLargeVolumetricDatasets}.
This pipeline implements a scalable algorithm to extract the annotated graph of a binary volume using a novel iterative refinement approach. \cref{fig:voreen_graph_samples} depicts two graphs generated by Voreen.

Based on these extracted graphs, we defined a selection of POIs on which to compute Saliency maps.
In vascular applications, three main relevant structural shapes can be identified: bifurcations (\emph{i.e.} vessel branching), vessel endings, and regular tubular branches. %
The first two are depicted by the nodes of the graphs, which correspond to specific logit in the image, while the latter is not directly linked to a specific logit but to a set of logits forming the centerlines. 
In order to also study the attribution on these regular tubular vessel, we defined the midpoints of the centerlines as POIs.

It is worth noting that we train our models following a 3D patch strategy with 25\% overlapping.
A single POI can therefore belong up to 8 patches.
In this case, we obtain as many Saliency maps as patches involved in the segmentation of the considered POI.

\subsection{Attribution}
Attribution maps are used to identify the input features (\emph{i.e.} voxels) that influence the model's decision the most.
Among the multiple approaches proposed over the past few years, gradient-based methods are a popular and intuitive way to produce attribution \cite{Ancona_2018_TowardsBetterUnderstandingOfGradientBasedAttributionMethods}. 
Considering a trained model as a non-linear function $F$, gradients represent the rate of change of a model's output $F(x)$ with respect to variations in the input $x$.

Despite the existence of more recent and sophisticated methods, we use Saliency maps \cite{Simonyan_2014_DeepInsideConvolutionalNetworksVisualisingImageClassificationModelsAndSaliencyMaps} as a baseline for our study.
Saliency is the simplest gradient-based method, where the signed gradient of the output with respect to the input is computed.
This can be mathematically expressed as:
\begin{equation}
    S_{i}(x)=\frac{\partial F(x)}{\partial x_{i}}
    \label{eq:saliency_deriv}
\end{equation}
where $x$ is the input image and $i$ is the $i^{th}$ voxel of $x$.

Saliency maps can involve both positive and negative values.
The sign of a voxel attribution indicates its influence on the logit prediction.
In our context of label-encoded binary segmentation, a positive voxel attribution corresponds to an increase in the output value of the target logit, thereby influencing the prediction towards the "vessel" class (label $1$).
Conversely, a negative gradient means that the voxel decreases the output value of the logit, thereby influencing the prediction towards the "background" class (label $0$).

\subsection{Analysis}
In this study, we aim to quantify the influence of domain-specific features on the model's decision-making process.
Domain-specific features include vessel and image characteristics.
Vessel characteristics cover the count of connected vessels (\emph{a.k.a.} connectivity), vessel thickness and volume, and the level of vessel tubularity.
Image characteristics cover the count and the total volume of vessel components in the patch.
As for Saliency maps, they reflect the model's behavior.
In particular, we measure descriptive statistics and L1-norm from the Saliency maps and relate them to domain-specific features.
The descriptive statistics are measured at global and local scale.

In order to investigate the model response to vascular patterns, we focus in this paper on the logits that have been finally classified as True Positive by the model.

\subsubsection{Relative Features in Patch}
    We based our model training on 3D patch data.
However, vessel characteristics have been determined considering the entire vascular structure.
As such, we would be comparing data from two different scales.
To mitigate this issue, we computed characteristics relatively to the patch scale.

\paragraph{Vessel Thickness} is estimated using the Exact Euclidean Distance Transform at the patches scale.

\paragraph{Tubularity Level} is determined by averaging the ouputs of several vesselness filters \cite{Lamy_2022_ABenchmarkFrameworkForMultiregionAnalysisOfVesselnessFilters}.
These filters are primarily based on the Hessian matrix and produce output values that can be interpreted as the probability of a voxel belonging to a tubular structure.

\paragraph{Vessel Connectivity} corresponds to the count of vessels connected to a given POI and was initally determined by the number of graph nodes connected to the considered POI.
Relative connectivity is the result of image processing involving the vascular skeleton.
First, we dynamically define an exclusion mask $M_{exc}$ that determines the minimum area to consider in order to capture the entire vascular pattern.
The size of this area is parametrized dynamically by extending the radius of a sphere, from $2$ to $10$ voxels.
At each radius iteration, we compute $card(NV)$ and $card(NB)$, where $NV$ and $NB$ represent the sets of new vessel voxels and new background voxels respectively.
When $card(NB)>0.75\times card(NV)$, we stop the expansion of the sphere and set the mask $M_{exc}$ based on its radius.

\begin{figure}[!ht]
    \centering
    \begin{subfigure}[b]{0.40\textwidth}
        \centering
        \includegraphics[width=\textwidth]{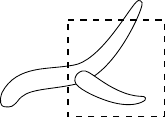}
        \caption{}
        \label{fig:relative_degree_1}
    \end{subfigure}
    \hfill
    \begin{subfigure}[b]{0.235\textwidth}
        \centering
        \includegraphics[width=\textwidth]{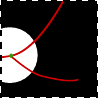}
        \caption{}
        \label{fig:relative_degree_2}
    \end{subfigure}
    \hfill
    \begin{subfigure}[b]{0.235\textwidth}
        \centering
        \includegraphics[width=\textwidth]{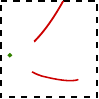}
        \caption{}
        \label{fig:relative_degree_3}
    \end{subfigure}
    \caption{Pipeline to determine the vessel connectivity degree relative to the patch. (\protect\subref{fig:relative_degree_1}) Patch extraction, which may distort the vessel structure. (\protect\subref{fig:relative_degree_2}) Exclusion mask $M_{exc}$ with the POI shown as a green dot and vessel centerlines as red curves. (\protect\subref{fig:relative_degree_3}) Resulting image $I_{cc}$ where only the vessels extending beyond $M_{exc}$ are retained to define relative connectivity.}
       \label{fig:relative_conectivity}
\end{figure}

We then consider the skeleton of vessels connected to the studied POI and identify those that extend beyond the exclusion area, as represented by $I_{cc}=I_{skel} \bigcap~!M_{exc}$.
Our new connectivity value, relative to the patch, is then the number of connected components in $I_{cc}$.
The \cref{fig:relative_conectivity} depicts a condensed diagram of our connectivity adjustment pipeline.
In this example, the connectivity of the vessel decreases from $3$ at the global scale (\cref{fig:relative_degree_1}) to $2$ at the patch scale (\cref{fig:relative_degree_3}).

\subsubsection{Local Analysis}
    A primary qualitative study of our Saliency maps shows that the most influential features are highly localized in a small compact region, while attribution values tend towards $0$ everywhere else.
In the following, we call these impactful areas \emph{blobs} because of their 3D blob-like shape. 

To better apprehend these highly influential regions, we have adopted a dual-scale analysis.
All attribution statistics extracted at the global scale — \emph{i.e.} considering the entire map — are also extracted in the blobs mask, at local scale.

\paragraph{Blob Detection} \label{sec:blobs_detection}
Popular blobs detectors such as LoG filter \cite{Lindeberg_1990_ScaleDetectionAndRegionExtractionFromAScaleSpacePrimalSketch} have limitations.
This latter searches for local maxima in intensity variations within images blurred by a range of Gaussian kernel sizes.
In numerous implementations, detected peaks are then interpreted as the centers of blobs, and the size of the blobs is set to the kernel size that gives the maximum response.
Hence, the output is limited to symmetric blobs, in a restricted size range.
In addition, this approach is sensitive and detects redundant and overlapping blobs, which is difficult to filter.

In our experiments, we opted for a custom blobs detector based on the multiscale Frangi filter \cite{Frangi_1998_MultiscaleVesselEnhancementFiltering}.
This filter is usually used to enhance the contrast of tube-like structures in an image such as vessels, but is also able to enhance blob-like structures.
The Frangi function (see \cref{eq:frangi}) relies on the second-order directional derivatives matrix, called the Hessian matrix $H$.
The study of the eigenvalues $\lambda_{i}$ of $H$ offers a way to enhance tube-like and blob-like structures.

\begin{equation}\label{eq:frangi}
    F = \begin{cases}
        \left(1-exp \left(-\frac{R_{a}^{2}}{2\alpha^{2}}\right) \right) exp \left(-\frac{R_{b}^{2}}{2\beta^{2}}\right) \left(1-exp \left(-\frac{S^{2}}{2C^{2}}\right) \right) & \text{if }\lambda_{2}, \lambda_{3} \leq 0\\
        0 & \text{otherwise}
    \end{cases}
\end{equation}

\begin{subequations}\label{eq:frangi_measures}
    \begin{gather}
        R_{b} = \frac{|\lambda_{1}|}{\sqrt{|\lambda_{2}\lambda_{3}|}}, \quad  
        R_{a} = \frac{|\lambda_{2}|}{|\lambda_{3}|}, \quad
        S=\|H\|=\sqrt{\lambda_{1}^{2}+\lambda_{2}^{2}+\lambda_{3}^{2}}
        \tag{\theequation a-c}
    \end{gather}
\end{subequations}

The parameters $\alpha$, $\beta$ and $C$ control sensitivity to deviation from plate-like, blob-like and high variance structures respectively.
As we lack ground-truth for the blobs, we determined the values of these hyperparameters empirically.
We retained the default parameters $\alpha=0.5, \beta=0.5$ and $C=15$ as they yielded satisfactory results during our qualitative evaluation.
Our Gaussian scale-space runs from $\sigma=2$ to $\sigma=16$, with a step fixed to $1$, covering a wide range of blob sizes.

The binary blobs mask is obtained after thresholding with Otsu's method, followed by the removal of small connected components.

The performance and results of our blob detector are sensitive to several key variables.
First, we must specify the type of blob to detect (white or black ridges).
The scale space parameter $\sigma$ is another significant parameter as it influences the size of the detected blobs.
Frangi's hyperparameters, $\alpha,~\beta$ and $C$ optimize the filter response for different structures shape.
The choice of thresholding method and related hyperparameters, like the number of bins for Otsu (here $n=256$), significantly impact the binary classification of pixels as part of a blob or background.
Finally, the size threshold for removing small connected components affects the final mask of detected blobs.

\begin{figure}[tb]
    \centering

    \begin{minipage}{0.32\textwidth}
        \centering
        \includegraphics[width=\textwidth]{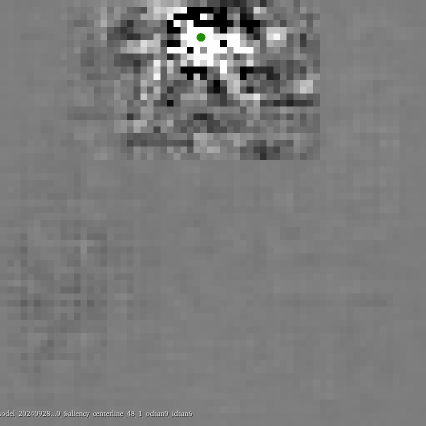}
        \label{fig:blobs_detection_attr}
    \end{minipage}
    \begin{minipage}{0.32\textwidth}
        \centering
        \includegraphics[width=\textwidth]{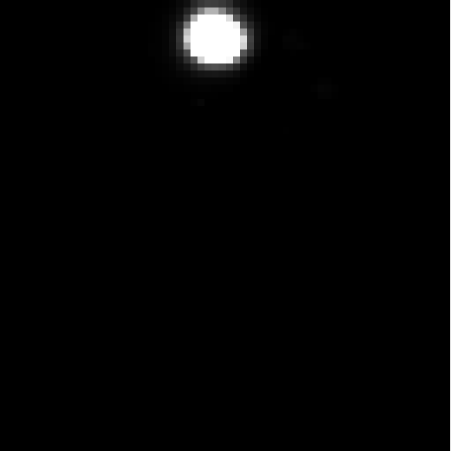}
        \label{fig:blobs_detection_frangi}
    \end{minipage}
    \begin{minipage}{0.32\textwidth}
        \centering
        \includegraphics[width=\textwidth]{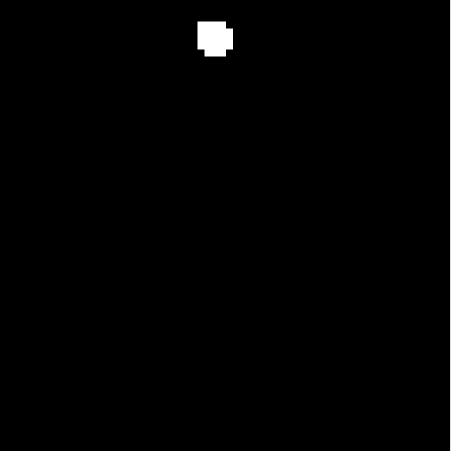}
        \label{fig:blobs_detection_mask}
    \end{minipage}

    \caption{Blob detection on the Saliency map. (Left) Saliency map where the green dot represents the explained POI. (Middle) Frangi filter output. (Right) Blob mask of positive attribution.}
    \label{fig:blob_detection}
\end{figure}

\section{Experiments}
    \subsection{Datasets}
    We conducted our experiments on two public vascular datasets: 3D-IRCADb-01 and Bullitt.

\paragraph{3D-IRCADb-01} \cite{Soler_2010_3dImageReconstructionForComparisonOfAlgorithmDatabase} (abbreviated as IRCAD) is a dataset of $20$ 3D CT-scans of the chest. 
Covering several thoracic organs, it has mainly been used to segment the liver and its vascular system over the past.

We allocated 5 images for the test set, from which vascular graphs extraction (see \cref{sec:vascularPOIs}) yielded $2,501$ vascular POIs.

In the context of explainability, this dataset is particularly valuable because 75\% of the patients have tumors.
This allows us to assess whether the attributions highlight behavioral changes in accordance with clinical reality.
Furthermore, IRCAD is one of the few public datasets focusing on the vascular system of the liver, an organ characterized by high heterogeneity in shape, size, and vasculature structure, making it a good candidate for our study.

However, the dataset has one major limitation with regard to annotations \cite{Lamy_2022_ABenchmarkFrameworkForMultiregionAnalysisOfVesselnessFilters}.
This is critical in our explainability context, as misclassified voxels distort our attribution analysis, which relies partly on the inference status (TP).
To mitigate this issue, we have extended our work using Bullitt as a second dataset.

\paragraph{Bullitt} \cite{Bullitt_2005_VesselTortuosityAndBrainTumorMalignancy} is a cerebral dataset of 100 MRA acquired from healthy volunteers.
Initially proposed as part of the TubeTK project, a subset of the original annotations (pairs of centerlines and diameters) has been refined, resulting in 33 voxel-wise annotated volumes \cite{Lamy_2021_VesselnessFiltersASurveyWithBenchmarksAppliedToLiverImaging}.

Similar to the IRCAD dataset, we reserved 5 images for the test phase.
Vessel graph extraction from the ground-truth yielded $6,243$ vascular POIs.

While the Bullitt dataset benefits from precise annotations and a larger number of data, enhancing the statistical reliability of our experiments, it also suffers from annotation limitations. Specifically, large veins in the peripheral region of the brain are not annotated.

\subsection{Models}
    We opted for a U-Net architecture due to its enduring status as a state-of-the-art model in vascular segmentation.
U-Net and its derivatives remain widely used and serve as a baseline for comparison in many recent publications \cite{Yan_2021_AttentionGuidedDeepNeuralNetworkWithMultiScaleFeatureFusionForLiverVesselSegmentation, Dang_2022_VesselCaptchaAnEfficientLearningFrameworkForVesselAnnotationAndSegmentation, Yu_2019_LiverVesselsSegmentationBasedOn3DResidualUNet}.
The U-Net architecture is effective and yet remains basic, simplifying our work compared to more complex models like nnU-Net \cite{Isensee_2021_nnUNetASelfConfiguringMethodForDeepLearningBasedBiomedicalImageSegmentation}, where inference relies on models ensembling.

We trained our models following the hyperparameters given in the work of Garret \emph{et al.} \cite{Garret_2024_DeepVesselSegmentationBasedOnANewCombinationOfVesselnessFilters}.
Unlike the latter, our models were trained from raw images, without any vesselness filter.

\section{Results}
    In our experiments, due to the patch strategy, a single input/output voxel can belong to up to 8 3D patches.
We remind that we focus solely on True Positive POIs, noting that the same spatial location may have varying prediction statuses across patches.
The results presented below are based on an analysis of $2,612$ Saliency maps for IRCAD and $8,529$ for Bullitt.

\subsection{Attribution Patterns and Blob Analysis}
    In nearly all cases --- $99.7 \%$ for IRCAD and $94.6\%$ for Bullitt --- the Saliency maps display highly localized, blob-shaped regions of positive attribution near the POI.
These are automatically detected using the blob detection method introduced in \cref{sec:blobs_detection}.

\cref{fig:blob_detection} illustrates an example of this process: the Saliency map (left) reveals high gradients near the POI, which are enhanced by the Frangi filter (middle) and then isolated into a binary blob mask (right).
This confirms the presence of compact, highly informative regions around the point being explained.

We assessed the contribution of attribution blobs by comparing them to the surrounding background in terms of information content.
First, we measured their contrast using the Fisher contrast-to-noise ratio \cite{Moreau_ContrastQualityControlForSegmentationTaskBasedOnDeepLearningModelsApplicationToStrokeLesionInCTImaging}, defined as:

\begin{equation}\label{eq:fisher_ctnr} F_{cnr} = \frac{(\mu_{blob}-\mu_{bg})^2}{\sigma^2_{blob} + \sigma^2_{bg}} \end{equation}

Here, $\mu_{blob}$ and $\mu_{bg}$ are the mean attribution values within the blob and background, respectively, and $\sigma_{blob}$, $\sigma_{bg}$ their standard deviations.
We then computed the average L1-norm of attribution values in both regions, and derived their ratio to estimate how much more influence the blob region has in the decision process.
Together, these metrics quantify the distinctiveness and relative importance of blob regions in the model’s output.

\begin{figure}[!ht]
    \centering
    \begin{subfigure}[b]{0.44\textwidth}
        \centering
        \includegraphics[width=\textwidth]{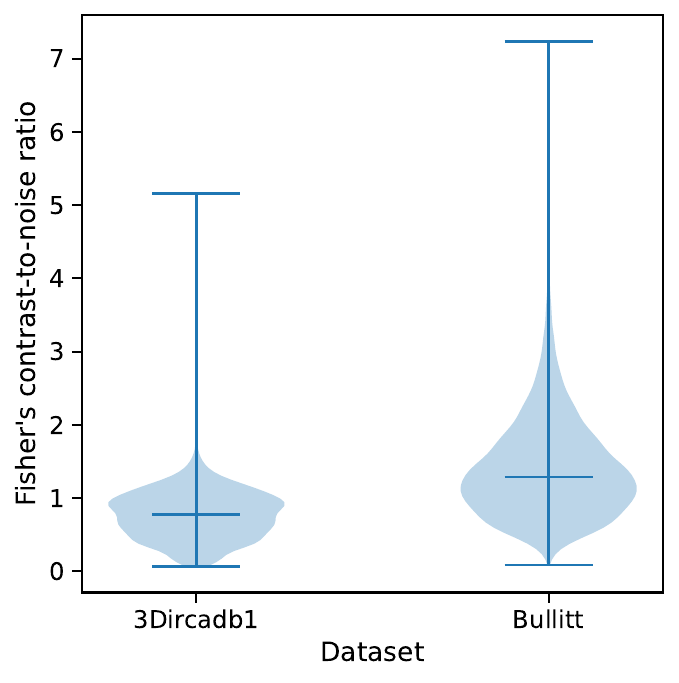}
        \caption{}
        \label{fig:fisher}
    \end{subfigure}
    \hfill
    \begin{subfigure}[b]{0.47\textwidth}
        \centering
        \includegraphics[width=\textwidth]{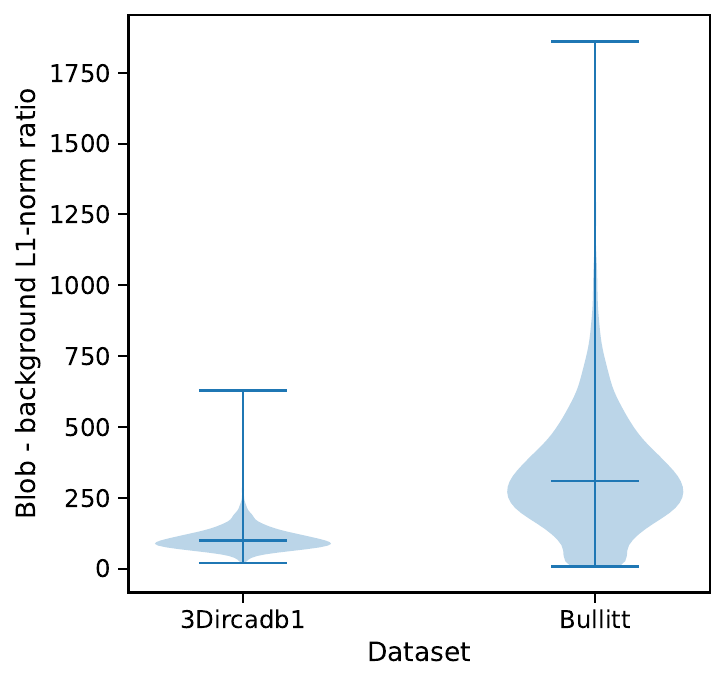}
        \caption{}
        \label{fig:l1_norm_ratio}
    \end{subfigure}
    \caption{Violin plot of (a) the Fisher contrast-to-noise ratio and (b) the average L1-norm ratio between blob attribution and background attribution, depending on the dataset.}
    \label{fig:contrast_measures}
\end{figure}

As shown in \cref{fig:contrast_measures} and \cref{table:fisher_results}, these metrics confirm that most of the explanatory power in the Saliency maps is concentrated in a very localized region.
For instance, in IRCAD, the average L1-norm inside the blob is $109 \pm 53.5$ times higher than outside, and even higher for Bullitt ($356 \pm 199$), highlighting the model’s strong local focus.

\begin{table}[ht]
    \small
    \begin{center}
    \caption{Attribution blobs contrast measures. \label{table:fisher_results}}
    \begin{tabular}{c | c c}
        & IRCAD & Bullitt \\
        \hline
        $F_{cnr}$ & $0.777 \pm 0.349$ & $1.45 \pm 0.708$ \\
        L1-norm ratio & $109 \pm 53.5$ & $356 \pm 199$
    \end{tabular}
    \end{center}
\end{table}

\subsection{Spatial Proximity of Attribution Blobs to POIs}
\begin{figure}
    \centering
    \includegraphics[width=0.47\textwidth]{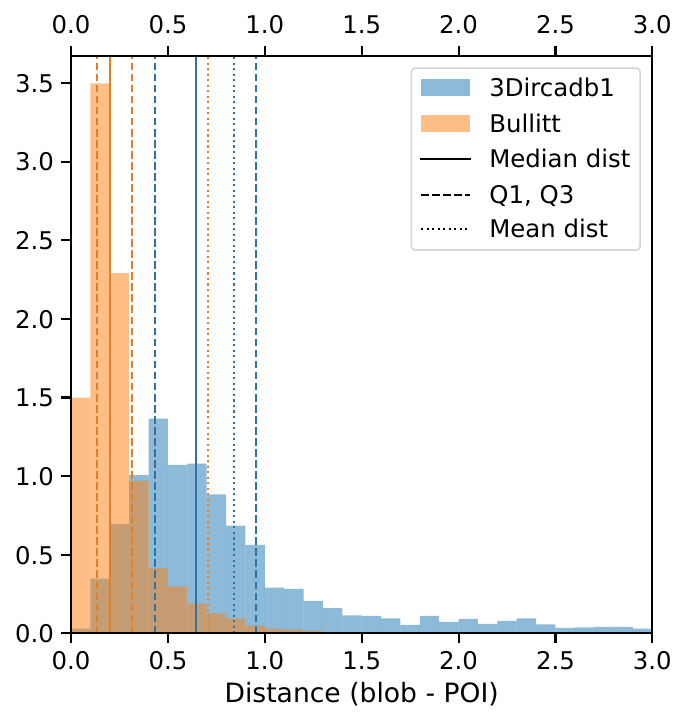}
    \caption{Histograms of distances between detected blobs and POIs.}
    \label{fig:dist_hist}
\end{figure}

    To further investigate the spatial behavior of the model, we computed the Euclidean distance between each detected blob centroid and its associated POI.
As illustrated in \cref{fig:dist_hist}, the distribution is heavily skewed toward low distances, with a median of $0.64$ voxel for IRCAD and $0.20$ for Bullitt.
The third quartile remains below $1$ voxel in both datasets, confirming that the model’s most influential features are located in the immediate neighborhood of the predicted point.

We observe that a few blobs are detected at larger distances from the POI, suggesting that some Saliency maps include multiple focus regions, not all centered around the prediction point.
These additional blobs may reflect either useful contextual reasoning --- \emph{e.g.}, attention to nearby vascular patterns --- or spurious responses to irrelevant features.
Differentiating these behaviors would require further analysis of multi-blob cases.

Extending the analysis to False Positives and False Negatives could reveal whether blob distance or multiplicity relates to prediction errors.
Similarly, examining True Negative POIs would clarify whether the model attends to vascular content or simply reacts to the queried locations.

\subsection{Domain-Specific Features Influence}
    We then analyzed the extent to which domain-specific vascular features --- such as vessel thickness, connectivity, or tubularity --- influence the model’s attributions.
We computed Spearman’s rank correlations between these features and various attribution metrics (norm, contrast, percentiles, and deviation).
The Spearman's rank correlation coefficient \cref{eq:spearman_s_correlation} allows us to identify dependencies between two variables described as monotonic functions.

\begin{equation}
    r_{s} = \frac{cov(rg_{x}, rg_{y})}{\sigma_{rg_{x}}\sigma_{rg_{y}}}
    \label{eq:spearman_s_correlation}
\end{equation}

\begin{figure}[!ht]
    \centering
    \includegraphics[width=\textwidth]{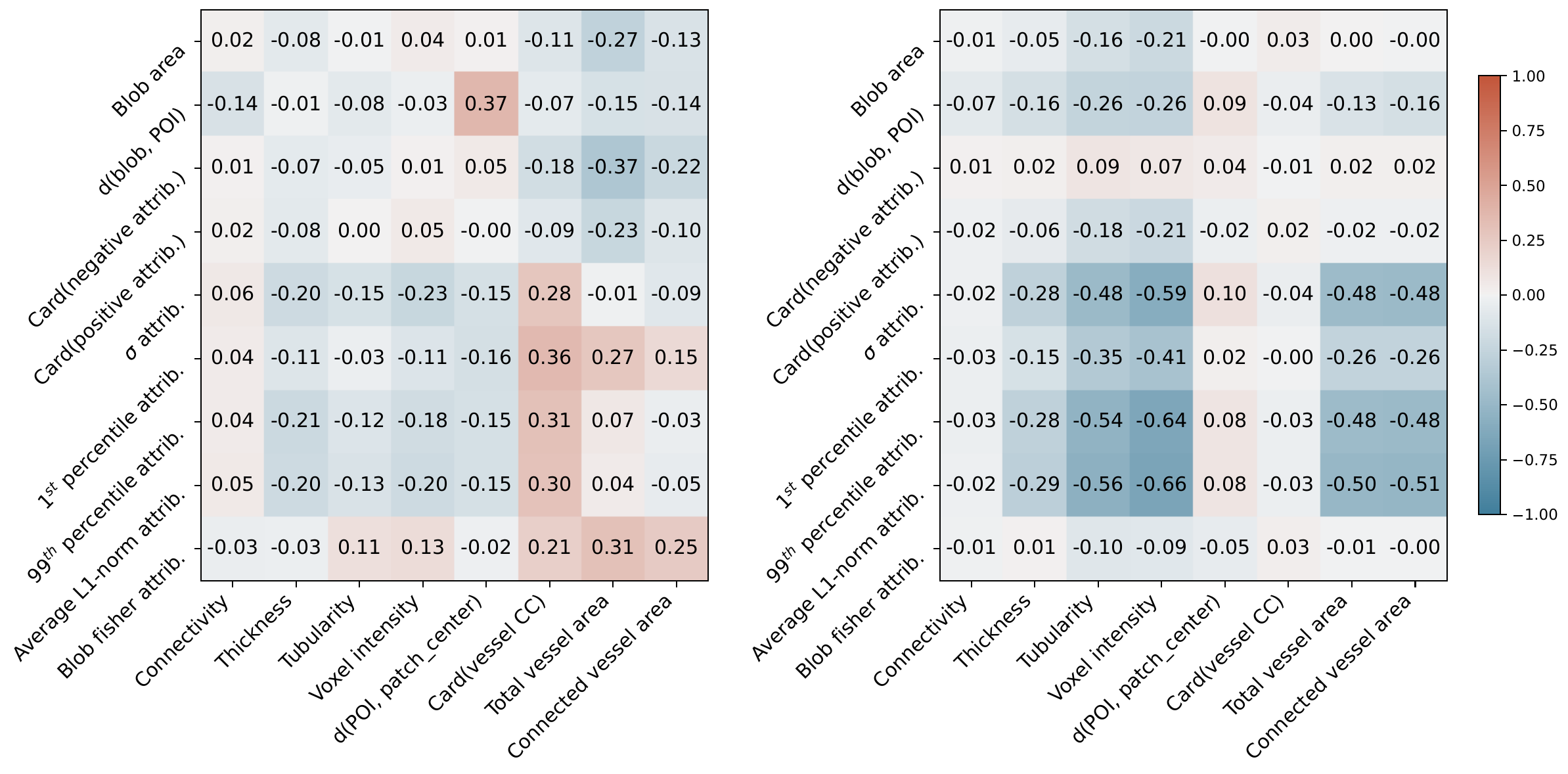}
    \caption{Spearman's rank correlations for IRCAD's (left) and Bullitt's (right) Saliency maps.}
    \label{fig:correlation_matrices}
\end{figure}

The results, shown in \cref{fig:correlation_matrices} , reveal generally weak correlations in the IRCAD dataset, and slightly stronger but still modest correlations in Bullitt.
In particular, tubularity appears negatively correlated with attribution strength in Bullitt, which is counterintuitive, as one might expect clearer, tubular structures to result in stronger positive attributions.

Overall, this suggests that the model relies predominantly on local visual cues rather than integrating higher-level vascular structure information in its decision-making process.

These findings suggest that the model’s attributions are driven primarily by local visual patterns, with little evidence that high-level anatomical features such as vessel connectivity, thickness, or tubularity are meaningfully integrated.
This indicates that the model does not capture the global vascular structure, contrary to what would be expected for robust and anatomically informed segmentation.
This misalignment calls for deeper investigation into the factors shaping attribution, and raises important questions about the model’s ability to generalize beyond local appearance cues.

\subsection{Attribution Sensitivity to Global Dataset Specificity}
\begin{figure*}[!ht]
    \centering
    \begin{subfigure}[h]{0.4\textwidth}
        \centering
        \includegraphics[width=\textwidth]{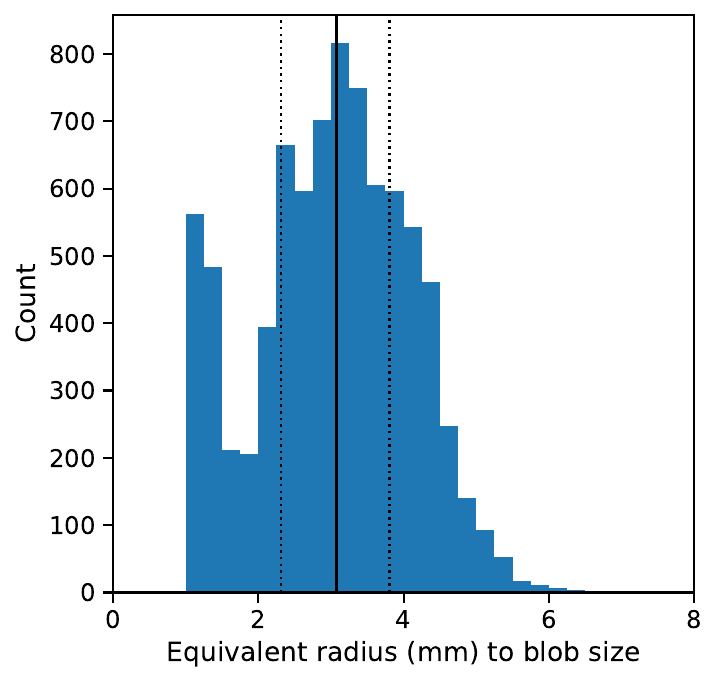}
        \caption[]{\small} 
        \label{fig:hist_ircad_blob_sizes}
    \end{subfigure}
    \begin{subfigure}[h]{0.416\textwidth}  
        \centering 
        \includegraphics[width=\textwidth]{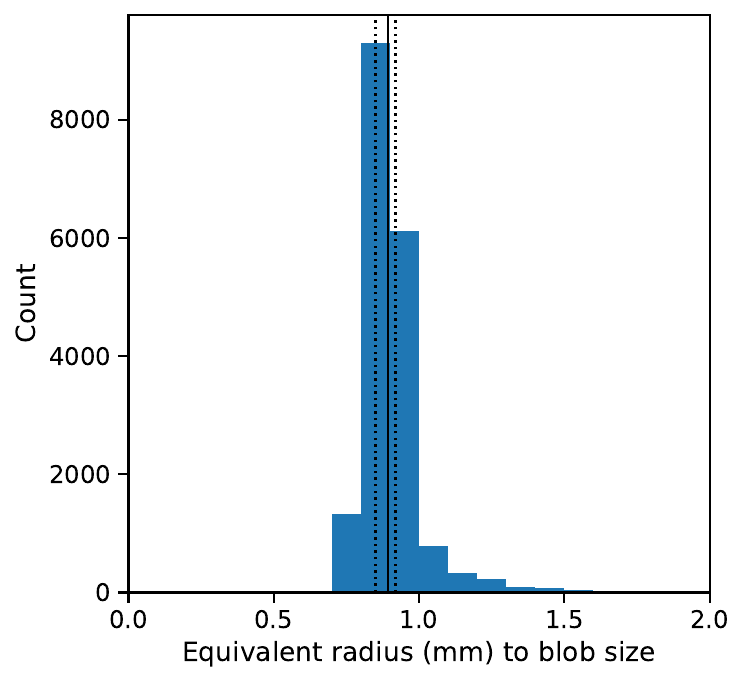}
        \caption[]{\small} 
        \label{fig:hist_bullitt_blob_sizes}
    \end{subfigure}
    \vskip\baselineskip
    \vspace{-.33cm}
    \begin{subfigure}[h]{0.4\textwidth}   
        \centering 
        \includegraphics[width=\textwidth]{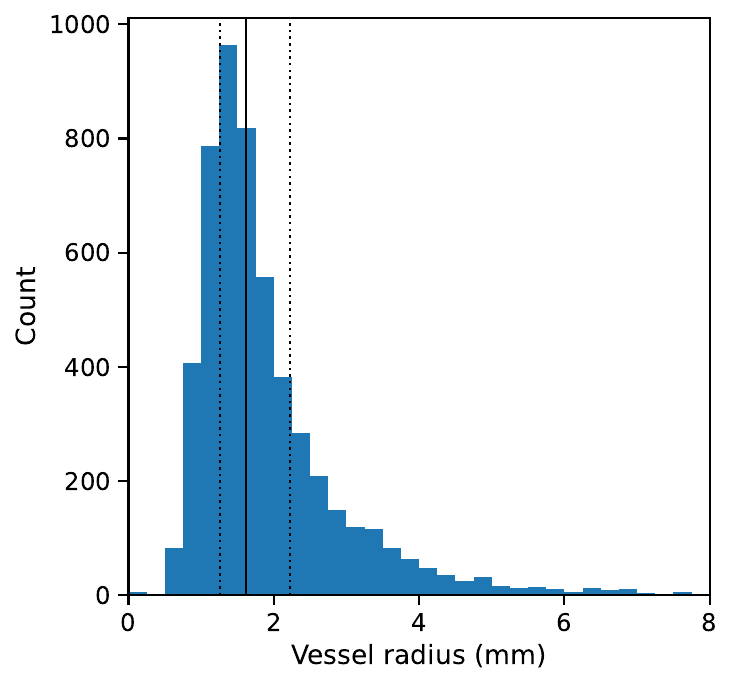}
        \caption[]{\small} 
        \label{fig:hist_ircad_vessels_size}
    \end{subfigure}
    \begin{subfigure}[h]{0.416\textwidth}   
        \centering 
        \includegraphics[width=\textwidth]{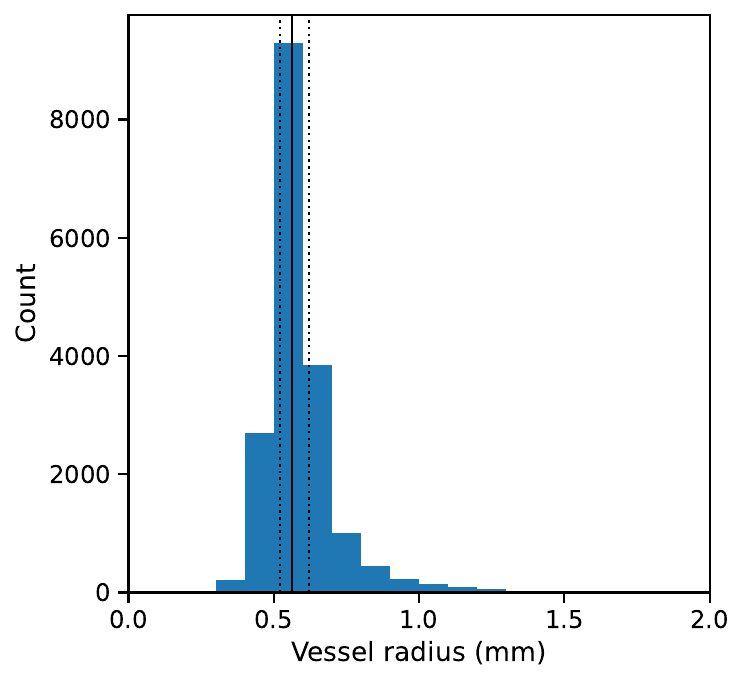}  
        \caption[]{\small} 
        \label{fig:hist_bullitt_vessels_size}
    \end{subfigure}

    \caption[]{\small Blobs size (top) and vessels size (bottom) distribution for IRCAD (left) and Bullitt (right) datasets.} 
    \label{fig:vessels_blobs_size_distributions}
\end{figure*}

    We observed systematic differences in blob sizes and variability across datasets.
As shown in \cref{fig:vessels_blobs_size_distributions}, IRCAD's Saliency maps typically feature a single large blob per map ($0.998 \pm 0.0552$ blob per map) with high size variability (\cref{fig:hist_ircad_blob_sizes}), while Bullitt's maps usually have a single blob but occasionally multiple ($0.991 \pm 0.646$ blob per map), which are smaller and less variable (\cref{fig:hist_bullitt_blob_sizes}).

Since attribution values do not correlate with domain-specific features at the patch scale, we assume that global dataset characteristics drive these differences.
One explanation could be image noise, which can hinder the model's focus, leading to larger and more variable blobs.
This aligns with our data, as IRCAD images are noisier ($PSNR=9.35 \pm 1.22$) than Bullitt ($PSNR=22.18 \pm 0.52$).

Another potential explanation lies in vessel size distribution.
We hypothesize that the model implicitly learns an "average vessel" scale during training, which may shape its effective field of view and influence the size of attribution blobs.
This is supported by the observed vessel size distributions: as shown in \cref{fig:hist_bullitt_vessels_size}, Bullitt vessels are smaller and more homogeneous, consistent with the smaller and less variable blobs in \cref{fig:hist_bullitt_blob_sizes}.
In contrast, \cref{fig:hist_ircad_vessels_size} shows that IRCAD vessels are larger and more heterogeneous, aligning with the broader blob size distribution in \cref{fig:hist_ircad_blob_sizes}.
While these patterns support the idea of scale adaptation, the correspondence between blob size and vessel size is not perfect, and further analysis is needed to confirm this relationship.

\section{Conclusion}
    In this work, we introduced a novel XAI pipeline to interpret vessel segmentation models by combining graph-based point selection and blob-level analysis of Saliency maps.
Our framework enables a fine-grained, anatomically informed evaluation of the model’s internal decision patterns.

Our results show that the model’s predictions are primarily driven by local visual cues in the immediate vicinity of each point of interest.
Despite the structural complexity of vascular networks, we found little evidence that the model leverages topological features such as connectivity, bifurcation patterns, or vessel orientation --- features that are critical for clinical reasoning.

Interestingly, we observed that blob sizes tend to reflect the average vessel scale specific to each dataset, suggesting that the model does adapt to simple global statistics derived from the training data.
This reveals an important discrepancy: while the model captures domain-specific cues at a coarse, statistical level (e.g., vessel size), it fails to integrate mid-level anatomical structure, which would be required for a more holistic understanding of vascular anatomy.

This gap highlights the limitations of standard convolutional architectures trained with voxel-wise supervision, which may perform well locally but overlook the broader anatomical context.
Future work should investigate the integration of topological priors, graph-based reasoning, or structural constraints to encourage models to move beyond local texture recognition and toward a more structured, anatomically aligned representation of vascular systems.

\section{Acknowledgments}
    This project receives support from the Région Auvergne-Rhône-Alpes under the DAISIES project. 
We thank the NVIDIA Academic Hardware Grant Program for providing the GPU resources that have been instrumental in carrying out the experiments for this work.

\section{Compliance with Ethical Standards}
    This research study was conducted retrospectively using human subject data made available in open access by the IRCAD Institute and TubeTK project.
Ethical approval was not required as confirmed by the license attached with the open access data.

\section{Disclosure of Interests}
    The authors have no competing interests to declare that are relevant to the content of this article.

\printbibliography %

\end{document}